\documentstyle[12pt,epsfig]{article}

\begin{document}
\baselineskip=23pt

\vspace{1.2cm}

\begin{center}
{\huge \bf   Unification of Gravitation, Gauge Field and Dark
Energy}

\bigskip

Xin-Bing Huang\footnote{huangxb@pku.edu.cn, huangxb@mail.phy.pku.edu.cn}\\
{\em Department of Physics,
Peking University,} \\
{\em  100871 Beijing, China}
\end{center}

\bigskip
\bigskip
\bigskip

\centerline{\large Abstract} This paper is composed of two
correlated topics: 1. unification of gravitation with gauge
fields; 2. the coupling between the daor field and other fields
and the origin of dark energy. After introducing the concept of
``daor field" and discussing the daor geometry, we indicate that
the complex daor field has two kinds of symmetry transformations.
Hence the gravitation and SU(1,3) gauge field are unified under
the framework of the complex connection. We propose a first-order
nonlinear coupling equation of the daor field, which includes the
coupling between the daor field and SU(1,3) gauge field and the
coupling between the daor field and the curvature, and from which
Einstein's gravitational equation can be deduced. The cosmological
observations imply that dark energy cannot be zero, and which will
dominate the doom of our Universe. The real part of the daor field
self-coupling equation can be regarded as Einstein's equation
endowed with the cosmological constant. It shows that dark energy
originates from the self-coupling of the space-time curvature, and
the energy-momentum tensor is proportional to the square of
coupling constant $\lambda$. The dark energy density given by our
scenario is in agreement with astronomical observations.
Furthermore, the Newtonian gravitational constant $G$ and the
coupling constant $\epsilon$ of gauge field satisfy $G=
\lambda^{2}\epsilon^{2}$.

\vspace{0.6 cm}

 PACS numbers: 95.35.+d, 04.60.Pp, 04.20.Cv, 02.40.Tt

 Keywords: the daor field, SU(1,3) gauge field, unification, dark
 energy

\vspace{1.2cm}

\newpage

\section{ Introduction\label{sec:Sec1}}
Recent astronomical observations on Type Ia
supernovae~\cite{sch98,rie98,per99} and the cosmological microwave
background radiation~\cite{ber00,jaf01,ben03} indicate that our
universe is spatially flat and accelerating at present, which
supports for a concordant cosmological model of inflation $+$ cold
dark matter $+$ dark energy. Furthermore all updated cosmological
observations do not conflict with a group of cosmological
parameters, namely $(\Omega_{\Lambda},\Omega_{M})
\approx(0.7,0.3)$~\cite{tag03}, which implies that the doom of our
universe will be dominated by dark energy~\cite{ft03}. The
properties of dark energy and its equation of state~\cite{mmot03}
have evoked much controversy in both astronomy and particle
physics communities~\cite{pr03,hua02}. In this paper we find that
dark energy originates from the self-coupling of the space-time
curvature. Firstly we will explain why we must introduce the daor
field in trying to unify gravitation with other interactions.

In the history there have been attempts to unify gravitation with
other interactions, notably the Kaluza-Klein approach of
compactifying higher dimensional space, and the
Einstein-Strauss-Schr$\rm \ddot{o}$dinger~\cite{ein46,sch85}
approach of considering a Hermitian metric tensor and interpreting
the antisymmetric field as that of the Maxwell field strength. The
Kaluza-Klein approach has been widely developed into modern
extra-dimensional theories, e.g., ADD scenario~\cite{add98} and
Randall-Sundrum models~\cite{rs99}, while the scenarios of the
complex space-time metric are less familiar~\cite{mof95}. It is
now well known that the antisymmetric part of the Hermitian metric
can not be explained as the electromagnetic field strength but
rather as an antisymmetric tensor where the theory is consistent
only if the field is massive~\cite{ddm93}. Following the method of
complexified geometry, Penrose proposed the twistor
theory~\cite{pen62} to study the quantization of gravitation in
1970s. In modern physics there are mainly two popular paradigms on
the unification of gravitation and quantum theory, that is, the
string theory~\cite{pol98} and the loop quantum
gravity~\cite{tal01}. The string theory is constructed on a
10-dimensional space-time (11-dimensional space-time for M
theory). Therefore there are enough extra dimensions to
accommodate the gauge fields. The most challengeable problem in
the string theory is how to compactify the extra dimensions to
give observable predictions. The loop quantum gravity theory
originates from Ashtekar's re-expression~\cite{ash86} of general
relativity, and inherits the geometric viewpoint from general
relativity. The loop quantum gravity has yielded several
interesting results, such as the discrete spectrum of the volume
operator.

Half a century ago, Yang and Mills constructed the non-Abelian
gauge field theory~\cite{yan54}. Based on the Yang-Mills theory
and the Higgs mechanism~\cite{hig64}, Glashow, Salam and Weinberg
set up a renormalizable electroweak gauge theory~\cite{gsw60}.
Combining this theory with quantum chromodynamics yields the
so-called $SU(3)_{C}\times SU(2)_{L}\times U(1)_{Y}$ standard
model of particle physics. Recently, it is claimed that a great
unified theory (GUT) endowed with SO(10) gauge group can give the
neutrino masses and mixing~\cite{xin04} which are not
contradictory to modern experiments~\cite{yan04}. If physicists
believe that the standard model or GUT should be a low-energy
approximation of a high energy unified quantum theory which
incorporate gravitation with gauge fields, then those inner
symmetry such as $SU(3)_{C}$, $U(1)_{Y}$ should be reasonably
accommodated in a higher energy theory. In this paper, we propose
a possibility to give those inner symmetries without introducing
extra-dimensions. Hence in our framework there is no difficulty on
compactification. After introducing the complex ``doar field" and
discussing the daor geometry, we obtain a complex spin connection
in which the space-time connection and SU(1,3) gauge field are
combined together. In the paper~\cite{hua04a}, we have argued that
quantizing gravitation needs to reformulate Einstein's equation
such that the new formalism must at least have the properties of a
complex field and the first-order differential equation. Therefore
we propose a first-order nonlinear equation for daor field, and
prove that Einstein's equation can be deduced from it. The real
part of the daor field self-coupling equation is equivalent to
Einstein's equation with the cosmological constant term. It shows
that dark energy originates from the self-coupling of the
space-time curvature.

The paper is organized as follows. In section 2 basic principles
are given, from which the complex daor field appears naturally.
The following section is devoted to the discussion of the doar
geometry, where the complex connection is acquired. In section 4
the unification of gravitation with SU(1,3) gauge field is
implemented, and a first-order nonlinear field equation is given,
from which Einstein's equation can be deduced. The origination of
dark energy is discussed in section 5. Conclusions and final
remarks are presented in the final section 6.

\section{Basic Principles\label{sec:Sec2}}

\renewcommand{\theequation}{2.\arabic{equation}}
\setcounter{equation}{0}

In this paper, let us suppose an ideal universe, in which there
are no fermion fields. Einstein's gravitational equation with the
cosmological constant term can be written as
 \footnote{In this paper, we use the
summation convention of Einstein: if an index occurs as both a
subscript and superscript in the same term, then the term is
summed over the range of the repeated index, and the summation
sign is omitted. Using Roman suffixes to refer to the bases of
local Minkowski frame; using Greek suffixes to refer to
curvilinear coordinates of space-time. }\cite{mtw73}
\begin{equation}
\label{daor1} R_{\mu\nu}-\frac{1}{2} ~ g_{\mu\nu}R=8\pi G
T_{\mu\nu}-\Lambda g_{\mu\nu}~,
\end{equation}
where $G$, $\Lambda$ are the Newtonian gravitational constant and
the cosmological constant respectively, and $T_{\mu\nu}$ is the
total energy-momentum tensor, in which the energy-momentum tensor
of gauge fields is denoted by $T^{\em{g}}_{\mu\nu}$.
 In the case of
electromagnetic field, $T^{\em{em}}_{\mu\nu}$ is given by
\begin{equation}
\label{daor101} 4\pi T^{\em{em}}_{\mu\nu}=\frac{1}{4} ~
g_{\mu\nu}{\bf f}_{\alpha\beta}{\bf f}^{\alpha\beta}-{\bf
f}_{\mu}^{~\alpha} {\bf f}_{\nu\alpha} ~,
\end{equation}
where ${\bf f}_{\alpha\beta}$ is the strength of electromagnetic
field. The sign convention adopted in this paper is the same as in
Misner-Thorne-Wheeler's book\cite{mtw73}. The metric tensor of
Minkowski space-time $\eta_{ab}$ is written as follows
\begin{eqnarray}\label{daor102}
\eta^{00}=-1~,~~\eta^{11}=\eta^{22}=\eta^{33}=+1~,~~
 \eta^{ab}=0~~{\rm for}~~a \neq b~.
\end{eqnarray}
We have proposed the following two basic principles in the
paper~\cite{hua04a}:

1. Physical space-time is a $3+1$ manifold, which looks like a
Minkowski space-time around each point.

2. The intrinsic distance
\begin{equation}
\label{daor3} ds^2= g_{\mu\nu}dx^{\mu}
dx^{\nu}~,~~~~~~\mu,\nu=0,1,2,3,
\end{equation}
keeps invariant under any physical transformations. One kind of
local transformations keeping $ds^2$ invariant is corresponding to
one kind of physical interactions.

In the literature, decomposing the metric into vierbeins or
tetrads $e^{a}_{~\mu}(x)$ has been adopted
extensively\cite{dir58,uti56,isr79}. The vierbein decomposition of
the space-time metric is expressed as
\begin{equation}
\label{daor301}
g_{\mu\nu}=\eta_{ab}e^{a}_{~\mu}(x)e^{b}_{~\nu}(x)~,
\end{equation}
which keeps $ds^{2}$ invariant under local SO$(1,3)$ group
transformation. But it is not the largest symmetry group as we
will show. Here we give the larger symmetry group transformations
which keep $ds^{2}$ invariant. To do so, we introduce the complex
vierbein (or tetrad)~\cite{cha04} field $k^{a}_{~\mu}$ or
$K^{~\mu}_{a}$, which satisfies
\begin{eqnarray}
\label{daor401}
2g_{\mu\nu}&=&\bar{k}^{~a}_{\mu}\eta_{ab}k^{b}_{~\nu}
+k^{~a}_{\mu}\eta_{ab}\bar{k}^{b}_{~\nu}~,
\\
\label{daor402}
2G^{\mu\nu}&=&{\bar{K}}^{\mu}_{~a}\eta^{ab}K^{~\nu}_{b}
+K^{\mu}_{~a}\eta^{ab}{\bar{K}}^{~\nu}_{b}~,
\\
\label{daor403}
g_{\mu\nu}G^{\nu\lambda}&=&G^{\lambda\nu}g_{\nu\mu}=\delta^{\lambda}_{\mu}~,
\end{eqnarray}
where bar denotes complex conjugation, the matrices $g$, $G$, $k$
and $K$ are all nonsingular. It is obvious that the matrix $g$ is
still real and symmetrical in Eq.(\ref{daor401}).

We have stressed in the introduction that the object studied in
this paper is the complex vierbein. To embody the linking property
of $k^{a}_{~\mu}$ between matter fields and the space-time
structure, also between gravitation and gauge interactions, here
we suggest giving the complex vierbein (or tetrad) field
$k^{a}_{~\mu}$ a new name, daor field\footnote{``Dao" is a basic
and important concept in ancient Chinese philosophy. ``Dao" is
used to refer not only the unobservable existence from which
everything originate but also the laws which dominate the doom of
everything. ``Dao" is also used to demonstrate the abstract
relationship between the dual things such as ``Yin" and ``Yang",
nihility and existence. Because $k^{a}_{~\mu}(x)$ plays such a
similar role in physics as will be shown in the following
sections, we suggest calling $k^{a}_{~\mu}(x)$ the ``daor
field".}.

As the space-time metric is real and symmetrical in general
relativity, there are at most 10 free-parameters in $g_{\mu\nu}$.
But Eq.(\ref{daor401}) and Eq.(\ref{daor402}) demonstrate that
there are at most 32 free-parameters in the daor field. It is too
much for describing the space-time metric from the viewpoint of
general relativity. To cancel nonphysical freedom, the following
covariant constraint should be satisfied
\begin{equation}
\label{daor-con}
\bar{k}^{~a}_{\mu}\eta_{ab}k^{b}_{~\nu}=k^{~a}_{\mu}\eta_{ab}\bar{k}^{b}_{~\nu}
~,~\bar{K}^{\mu}_{~a}\eta^{ab}K^{~\nu}_{b}
=K^{\mu}_{~a}\eta^{ab}\bar{K}^{~\nu}_{b} ~,
\end{equation}
then Eq.(\ref{daor401}) and Eq.(\ref{daor402}) become
\begin{equation}
\label{daor4} g_{\mu\nu}=\bar{k}^{~a}_{\mu}\eta_{ab}k^{b}_{~\nu}~
,~~~~~~G^{\mu\nu}=\bar{K}^{\mu}_{~a}\eta^{ab}K^{~\nu}_{b}~.
\end{equation}

Define a tensor $N_{\mu\nu}$ with respect to curvilinear
coordinate as follows
\begin{equation}
\label{zero-matrix}
2N_{\mu\nu}=\bar{k}^{~a}_{\mu}\eta_{ab}k^{b}_{~\nu}
-k^{~a}_{\mu}\eta_{ab}\bar{k}^{b}_{~\nu}~.
\end{equation}
The above equation shows that the tensor $N_{\mu\nu}$ is
antisymmetrical, namely $N_{\mu\nu}=-N_{\nu\mu}$. Thus the
covariant constraint $N_{\mu\nu}=0$ only provides 6 independent
constraint equations to the components of daor field. That is to
say, the daor field can at most have 26 independent
free-parameters.

Consider general real coordinate transformations $x\rightarrow
x^{\prime}(x)$, since
\begin{equation}
\label{co-trans} dx^{\prime\mu}=\frac{\partial x^{\prime\mu}
}{\partial x^{\nu}}dx^{\nu}~ ,~~~~~~k^{\prime
b}_{~~\mu}=k^{b}_{~\nu}\frac{\partial x^{\nu}}{\partial
x^{\prime\mu} }~,
\end{equation}
the intrinsic distance is invariant under general coordinate
transformations.

Under the rotation of the locally complexified Minkowski frame,
the daor field $k^{b}_{~\nu}$ transforms as follows
\begin{equation}
\label{ro-trans} k^{a}_{~\nu}\rightarrow k^{\prime
a}_{~\nu}=S^{a}_{~b}k^{b}_{~\nu}~.
\end{equation}
If the matrix $S^{a}_{~b}$ satisfies
\begin{equation}
\label{ro-trans01} \bar{S}^{~c}_{a}\eta_{cd}S^{d}_{~b}=\eta_{ab}~,
\end{equation}
namely, $S^{a}_{~b}$ being the element of SU(1,3) group, then the
intrinsic distance is invariant under the rotation of the locally
complexified Minkowski frame.

Hence we can draw a conclusion: by introducing the complex daor
field, we find that the intrinsic distance and the covariant
constraint are invariant under two kinds of transformations: one
is the general real coordinate transformation $x\rightarrow
x^{\prime}(x)$; the other is the SU(1,3) group transformation of
the Roman suffixes.

\section{Daor Geometry and Its Connection\label{sec:Sec3}}

\renewcommand{\theequation}{3.\arabic{equation}}
\setcounter{equation}{0}

Obviously the daor field $k^{a}_{~\mu}$ satisfying the constraint
Eq.(\ref{daor-con}) can be reexpressed in terms of the real
vierbeins as
\begin{equation}
\label{conn001} k^{a}_{~\mu}(x)=l^{a}_{~b}(x)e^{b}_{~\mu}(x)~,
\end{equation}
where $e^{b}_{~\mu}(x)$ is the real vierbein defined by
Eq.(\ref{daor301}), and the matrix $l^{a}_{~b}(x)$ satisfies
Eq.(\ref{ro-trans01}), namely, $l^{\dag}~\eta ~l=\eta$.

To provide the mathematical tools for an advanced study in the
daor field, we would like to introduce the main results of daor
geometry given by Ref.\cite{hua04a}. It can be defined
that\footnote{We use $\partial_{\mu}$ to denote the partial
differential operator $\frac{\partial}{\partial x^{\mu}}$ for
brevity.}
\begin{eqnarray}\label{conn002}
e^{a}=e^{a}_{~\mu}{\rm
d}x^{\mu}~,~~k^{a}=l^{a}_{~b}e^{b}~,~~k^{\dag a}=e^{b}
\bar{l}^{~a}_{b}~,
~~{K}^{\dag}_{a}=\bar{K}^{\mu}_{~a}\partial_{\mu}~,
~~K_{a}=K^{~\mu}_{a}\partial_{\mu}~,
\end{eqnarray}
where $k^{a}$ and $k^{\dag a}$ are the daor field 1-forms,
$\bar{K}^{\mu}_{~a}$ and $K^{~\mu}_{a}$ are given by
\begin{equation}
\label{conn003} {K}^{\dag}=k^{-1}~, ~~~~~~K^{-1}=k^{\dag}~.
\end{equation}
It is proved that the set $\{
{K}^{\dag}_{0},{K}^{\dag}_{1},{K}^{\dag}_{2},{K}^{\dag}_{3} \}$ is
the basis of locally complexified Minkowski frame ${\bf M}$ and
the set $\{ k^{0},k^{1},k^{2},k^{3} \}$ is a basis of its dual
frame ${\bf M}^{*}$. Hence a vector $U$ and a covector $V$ are
denoted by $U=U^{a}{K}^{\dag}_{a}$ and $V= V_{a} k^{a}$
respectively. Therefore an $(r,s)$-type tensor $T$ can be uniquely
expressed as
\begin{equation}
\label{tensor003} T=T^{{i}_1\cdots {i}_r}_{j_1 \cdots j_s}
{K}^{\dag}_{i_1} \otimes \cdots \otimes {K}^{\dag}_{i_r} \otimes
k^{j_{1}} \otimes \cdots \otimes k^{j_{s}}~,
\end{equation}
where $T^{{i_1}\cdots {i_r}}_{j_1 \cdots j_s}$ are called the
components of the tensor $T$.

The Cartan's exterior product and exterior differentiation can
also be built up in daor geometry. For a $p$-form
$\alpha_p=\frac{1}{p!}f_{a_1 \cdots a_p}k^{a_1}\wedge \cdots
\wedge k^{a_p}$, its exterior differentiation is defined as the
following
\begin{equation}
\label{diff003} {\rm d}\alpha_p=\frac{1}{p!}f_{a_1 \cdots
a_p,l}k^{l} \wedge k^{a_1}\wedge \cdots \wedge k^{a_p}~.
\end{equation}
Let $\alpha_p\in\Lambda^{p}({\bf M})$ and
$\beta_q\in\Lambda^{q}({\bf M})$, their exterior product satisfies
$\alpha_p\wedge
\beta_q=(-1)^{pq}\beta_q\wedge\alpha_p$~\cite{ccl99,hh97}.

For a covector field $\alpha_1=f_a (x) k^{a}(x)$, when the daor
field is chosen, then
\begin{equation}
\label{conn008} \nabla \alpha_1={\rm d} f_a\otimes k^{a}+f_a\nabla
k^{a} ~.
\end{equation}
The above equation shows that $\nabla \alpha_1$ can be calculated
if the covariant differentiation $\nabla k^{a}$ of the daor field
is given. $\nabla k^{a}$ denote the infinitesimal variance of the
daor field $k^a$ at the neighborhood of a point and can be
expressed as
\begin{eqnarray}\label{conn00444}
 \nabla{k}^{a}=(\nabla
l^{a}_{~b})e^{b}+l^{a}_{~b}(\nabla e^{b})
=-l^{a}_{~c}B^{c}_{~b}e^{b}-l^{a}_{~b}\theta^{b}_{~c}e^{c}
 =-l^{a}_{~b}\omega^{b}_{~c}e^{c}~,
\end{eqnarray}
where
\begin{equation}
\label{conn010}
\omega^{a}_{~b}=B^{a}_{~b}+\theta^{a}_{~b}=-<K^{\dag}_{b},\nabla
k^{a}
>=\omega^{a}_{ib}k^{i} ~.
\end{equation}
Since $\omega^{a}_{~b}$ is the complex matrix valued 1-form, we
suggest calling $\omega^{a}_{~b}$ the daor connection 1-form.

It is well known that $\theta^{a}_{~b}$ defined in
Eq.(\ref{conn00444}) is the spin connection introduced first by
Cartan. From the viewpoints of Yang-Mills gauge field, the daor
connection $\omega^{a}_{~b}$ is the SU(1,3)$\times$SO(1,3) gauge
field. Or equivalently, in the language of differential geometry,
we can say that $\omega^{a}_{~b}$ is the connection on
SU(1,3)$\times$SO(1,3) principal bundle. The curvature of this
principal bundle is thus expressed as
\begin{equation}
\label{daor1903} \Omega^{a}_{~b}={\rm d
}\omega^{a}_{~b}+\omega^{a}_{~c}\wedge\omega^{c}_{~b}~.
\end{equation}
Because of Eq.(\ref{conn010}), the curvature $\Omega^{a}_{~b}$ can
be written as
\begin{eqnarray}
\label{daor190301}
 \Omega^{a}_{~b}=
R^{a}_{~b}+F^{a}_{~b}+\theta^{a}_{~c}\wedge
B^{c}_{~b}+B^{a}_{~c}\wedge\theta^{c}_{~b}~,
\end{eqnarray}
where $R^{a}_{~b}$, $F^{a}_{~b}$ are the curvature of SO(1,3)
principal bundle and the curvature of SU(1,3) principal bundle
respectively. The definitions of $R^{a}_{~b}$ and $F^{a}_{~b}$ are
\begin{eqnarray}
\label{daor190303} R^{a}_{~b}&=&{\rm d
}\theta^{a}_{~b}+\theta^{a}_{~c}\wedge \theta^{c}_{~b}~,\\
\label{daor1902++} F^{a}_{~b}&=&{\rm d }B^{a}_{~b} + B^{a}_{~c}
\wedge B^{c}_{~b}~.
\end{eqnarray}

The equation (\ref{conn00444}) shows that there are two categories
of local gauge transformations: One is the local SU(1,3) group
transformation, the covariant principle naturally leads to a
necessary input of SU(1,3) gauge field; The other is the local
SO(1,3) group transformation. The spin connection
$\theta^{a}_{~b}$ represents the effects of gravitation.

Firstly, let us consider an intrinsic rotation of the daor field
\begin{equation}
\label{daor15} k^{a}\rightarrow k^{\prime a}=l^{\prime a
}_{~b}e^{b}=S^{a}_{~b}k^{b}~,
\end{equation}
where $S^{a}_{~b}$ satisfies Eq.(\ref{ro-trans01}), namely,
$S^{a}_{~b}$ is a faithful representation of SU(1,3) group.
$l^{\prime a }_{~b}$ is defined by $l^{\prime a
}_{~b}=S^{a}_{~c}l^{c}_{b}$. From Refs.\cite{hua04a,egf80}, it is
known that, under the intrinsic SU(1,3) rotation of the daor
field, the daor connection 1-form $\omega^{a}_{b}$ transforms as
follows
\begin{equation}
\label{daor17} \omega^{\prime
a}_{~~b}=S^{a}_{~c}\omega^{c}_{~d}(S^{-1})^{d}_{~b}+S^{a}_{~c}({\rm
d }S^{-1})^{c}_{~b}~,
\end{equation}
Since $B^{a}_{~b}$ is the connection of SU(1,3) principal bundle,
under the SU(1,3) gauge rotation of the daor field, $B^{a}_{~b}$
transforms into
\begin{equation}
\label{daor19+} B^{\prime
a}_{~~b}=S^{a}_{~c}B^{c}_{~d}(S^{-1})^{d}_{~b}+ S^{a}_{~c}({\rm d
}S^{-1})^{c}_{~b}~,
\end{equation}
and $\theta^{a}_{~b}$  satisfies
\begin{equation}
\label{daor1901} \theta^{\prime
a}_{~~b}=S^{a}_{~c}\theta^{c}_{~d}(S^{-1})^{d}_{~b}~.
\end{equation}
Furthermore, it is easy to prove that under this rotation
$F^{a}_{~b}$ and $\Omega^{a}_{~b}$ become
\begin{equation}
\label{daor190302} {F}^{\prime
a}_{~~b}=S^{a}_{~c}F^{c}_{~d}(S^{-1})^{d}_{~b}~,~~~~{\Omega}^{\prime
a}_{~~b}=S^{a}_{~c}\Omega^{c}_{~d}(S^{-1})^{d}_{~b}~.
\end{equation}

Secondly, consider an orthogonal rotation of the real orthonormal
vierbein
\begin{equation}
\label{daor190304} e^{a}\rightarrow e^{\prime a}=
\Phi^{a}_{~b}e^{b}~,
\end{equation}
where $\Phi^{a}_{~b}$ satisfies
\begin{equation}
\label{daor19030400}
\Phi^{~a}_{c}\eta_{ab}\Phi^{b}_{~d}=\eta_{cd}~.
\end{equation}
Eq.(\ref{daor19030400}) demonstrates that $\Phi^{a}_{~b}(x)$ is a
representation of SO(1,3) group. The doar field transforms as
\begin{equation}
\label{daor1903041} k^{a}\rightarrow k^{\prime a}=l^{\prime a
}_{~c}e^{\prime c}=\Phi^{a}_{~b}k^{b}~,
\end{equation}
where $l^{\prime a }_{~c}=\Phi^{a}_{~b}
l^{b}_{~e}(\Phi^{-1})^{e}_{~c}$. After this transformation, the
new daor connection is
\begin{equation}
\label{daor190305} \omega^{\prime
a}_{~~b}=\Phi^{a}_{~c}\omega^{c}_{~d}(\Phi^{-1})^{d}_{~b}+\Phi^{a}_{~c}({\rm
d }\Phi^{-1})^{c}_{~b}~,
\end{equation}
Similarly, we acquire
\begin{equation}
\label{daor190306} \theta^{\prime
a}_{~~b}=\Phi^{a}_{~c}\theta^{c}_{~d}(\Phi^{-1})^{d}_{~b}+
\Phi^{a}_{~c}({\rm d }\Phi^{-1})^{c}_{~b}~,
\end{equation}
and
\begin{equation}
\label{daor190307} B^{\prime
a}_{~~b}=\Phi^{a}_{~c}B^{c}_{~d}(\Phi^{-1})^{d}_{~b}~.
\end{equation}
The transformation formulae for the curvatures $R^{a}_{~b}$ and
$\Omega^{a}_{~b}$ are given by
\begin{equation}
\label{daor190308} R^{\prime
a}_{~~b}=\Phi^{a}_{~c}R^{c}_{~d}(\Phi^{-1})^{d}_{~b}~,~~~~{\Omega}^{\prime
a}_{~~b}=\Phi^{a}_{~c}\Omega^{c}_{~d}(\Phi^{-1})^{d}_{~b}~.
\end{equation}


\section{Unification of Gravitation with Gauge Fields\label{sec:Sec5}}
\renewcommand{\theequation}{5.\arabic{equation}}
\setcounter{equation}{0}

Consider a kind of gauge groups, which are the subgroups of
SU(1,3) group, and the element of which can be written as
\begin{equation}
\label{daor7} Z_{~b}^{a}(x)=e^{it^{a}_{~b}(x)}~.
\end{equation}
Where $t^{a}_{~b}$ is a $4\times 4$ matrix, which is traceless and
Hermitian, and, of course, should be the function of curvilinear
coordinates. Therefore $Z^{a}_{~b}$ is a 4-dimensional
representation of gauge group. Some groups such as U(1), SU(2) all
satisfy the condition (\ref{daor7}).

In this paper we only discuss the SU(1,3) gauge field. Covariance
of the daor field equations under the local SU(1,3) group rotation
directly leads to the introduction of Yang-Mills gauge
field~\cite{yan54}, say, denoted by ${\tilde B}^{a}_{~bc}k^{c}$.
Comparing Eq.(\ref{daor19+}) with the formulae in Yang-Mills
theory, one obtains
\begin{eqnarray}
\label{daor18}  B^{a}_{~b}= i \epsilon {\tilde B}^{a}_{~b}~,~~
{\tilde B}^{\prime a}_{~~b}=Z^{a}_{~c}{\tilde
B}^{c}_{~d}(Z^{-1})^{d}_{~b}+ \frac{1}{i\epsilon}Z^{a}_{~c}({\rm d
}Z^{-1})^{c}_{~b}~,
\end{eqnarray}
where $\epsilon$ is the coupling constant of the SU(1,3) gauge
field. Both ${\tilde B}^{a}_{~b}$ and $B^{a}_{~b}$ are 1-forms.
The gauge field strengths ${\tilde F}^{a}_{~b}$ corresponding to
the gauge field ${\tilde B}^{a}_{~b}$ reads
\begin{equation}
\label{daor1902} {\tilde F}^{a}_{~b}={\rm d }{\tilde
B}^{a}_{~b}+i{\epsilon}{\tilde B}^{a}_{~c}\wedge {\tilde
B}^{c}_{~b}~.
\end{equation}
It is obvious that $F^{a}_{~b}=i \epsilon{\tilde F}^{a}_{~b}$.

The energy-momentum tensor of the SU(1,3) gauge field can be
written in the form~\cite{qui83,car82}
\begin{equation}
\label{daor20} 4\pi T^{\em{g}}_{\mu\nu}=\frac{1}{4} ~
g_{\mu\nu}{\rm tr}({\tilde F}_{\alpha\beta}{\tilde
F}^{\alpha\beta}) -{\rm tr}( {\tilde F}_{\mu}^{~\alpha} {\tilde
F}_{\nu\alpha})~,
\end{equation}
where `tr' means operation of acquiring the trace of a matrix, and
$T^{\em{g}}_{\mu\nu}$ is, as can easily be seen, traceless,
$g^{\mu\nu}T^{\em{g}}_{\mu\nu}=0$, but
$\nabla^{\mu}T^{\em{g}}_{\mu\nu}\neq 0$. In Eq.(\ref{daor20}), we
have introduced the physical gauge fields ${\bf
f}_{\alpha\beta}^{I},~I=1,2,\ldots,15$, as follows
\begin{equation}
\label{daor20+} {\tilde F}_{\alpha\beta}\equiv
\frac{1}{2}\sum_{I}{\bf T}^{I}{\bf f}_{\alpha\beta}^{I}~,
\end{equation}
where ${\bf T}^{I},~I=1,2,\ldots,15$, are fifteen $4\times4$
generators of SU(1,3) group. The Lie algebra of SU(1,3) group is
of the form
\begin{equation}
\label{daor2+} [{\bf T}^{I},{\bf T}^{J}]= i C^{IJ}_{K}{\bf
T}^{K}~,
\end{equation}
here $C^{IJ}_{K}$ being the structure constants of SU(1,3) group.
Furthmore, the generators ${\bf T}^{I}$'s are selected so that
${\bf T}^{I}$ satisfy
\begin{equation}
\label{daor20++} {\bf T}^{I}{\bf T}^{I}={\bf 1}~,
\end{equation}
where ${\bf 1}$ denotes $4\times4$ unit matrix. Since we only
consider the coupling between the daor field and the SU(1,3) gauge
field, the energy-momentum tensor in Einstein's equation
(\ref{daor1}) must include the term of Eq.(\ref{daor20}).

In the case of Einstein's gravitational theory, the Levi-Civita
connection $\Gamma^{\mu}_{\alpha\beta}$ is adopted. In Riemannian
geometry the Levi-Civita connection is determined by two
conditions, the covariant constancy of the metric and the absence
of torsion. In our case, the daor connection is complex. We need
not the torsion-free condition. In the paper~\cite{hua04a} we have
argued that quantizing gravitation needs to rebuild Einstein's
gravitational theory such that the new formalism at least includes
two properties: 1, the complexified field; 2, the first-order
nonlinear field equation. In this paper, we try to unify
gravitation with gauge fields and pave the way for quantizing this
theory. Therefore, the daor field equation should have the
following properties:
\begin{enumerate}
\item Einstein's general relativity is a special result of this
theory. \item Gauge fields are naturally introduced and
incorporated with gravitation. \item The equation is consistent
with the basic principles of quantum theory, e.g., the physical
operators appearing in the theory are Hermitian. \item The theory
can give some new predictions.
\end{enumerate}

Because of these considerations, we propose the fundamental doar
field equation as follows
\begin{equation}
\label{daor21} {\rm d }k^{a}+(\omega^{a}_{~b}+\lambda~{\cal
R}^{a}_{~b})\wedge~k^{b}=0~.
\end{equation}
Where $\lambda$ is the real coupling constant, which will be given
at the end of this section. Where ${\cal R}^{a}_{~b}$ is defined
by
\begin{eqnarray}\label{daor22}
{\cal R}^{a}_{~b}\equiv i_{X}\Omega^{a}_{~b}=<X,\Omega^{a}_{~b}> =
<\gamma^{c}K^{\dag}_{c},~\frac{1}{2}\Omega^{a}_{~bij}k^{i}\wedge
k^{j}> =\gamma^{c}\Omega^{a}_{~bci}k^{i}~.
\end{eqnarray}
Here $X=\gamma^{a}K^{\dag}_{a}$ is a vector field, $\gamma^{a}$
denotes the standard flat-space Dirac matrices, which satisfies
$\{\gamma^{a},\gamma^{b}\}=2\eta^{ab}$.

Now we would like to demonstrate the properties embodied by
Eq.(\ref{daor21}). First, Eq.(\ref{daor21}) is covariant under
both the local SU(1,3) gauge transformations and the local SO(1,3)
rotations of the moving frame. Secondly, the covariant constancy
of the metric makes sure that $\theta^{a}_{~b}=-\theta^{b}_{~a}$,
hence the operator $i{\rm d }+i(\omega^{a}_{~b}+\lambda~{\cal
R}^{a}_{~b})\wedge$ is Hermitian, which is consistent with the
requirement of quantum theory. Finally, in quantum field theories,
all gauge fields are introduced by the local gauge invariance of
the spinor fields. Therefore, in the daor field framework, the
spinor fields should be included also. Obviously the daor field
described by Eq.(\ref{conn001}) can be naturally extended to
\begin{equation}
\label{m001}
k^{a}_{~\mu}(x)=l^{a}_{~b}(x)e^{b}_{~\mu}(x)
~,~~~~~~~~l^{a}_{~b}(x)=[\psi(x)]^{a}_{~b}~,
\end{equation}
where $\psi(x)$'s are sixteen $4 \times 1$
matrices\footnote{Whether these $4 \times 1$ matrices can be
treated as the spinor fields will be studied in the forthcoming
paper.}, which are also constrained by
\begin{equation}
\label{m002}
[\psi^{\dag}(x)]^{~c}_{a}\eta_{cd}[\psi(x)]^{d}_{~b}=\eta_{ab}~.
\end{equation}
The above extension of the daor field implies that Dirac matrices
$\gamma_{a}$ may appear in the Eq.({\ref{daor21}}), or we can say,
should uniquely appear in the vector $X$. This point reasonably
demonstrates why the vector $X$ should be selected as
$\gamma^{a}K^{\dag}_{a}$.

The covariant exterior derivative of a (1,1)-type tensor valued
differential form $V^{a}_{~b}$ of degree $p$ is defined
as\cite{egf80}
\begin{equation}
\label{daor25} DV^{a}_{~b}={\rm d } V^{a}_{~b}+\omega^{a}_{~c}
\wedge V^{c}_{~b}- (-1)^{p} V^{a}_{~c} \wedge \omega^{c}_{~b}~.
\end{equation}
Exteriorly differentiating Eq.(\ref{daor1903}), we can find the
Bianchi identity
\begin{equation}
\label{bian01} {\rm d } \Omega^{a}_{~b}+\omega^{a}_{~c} \wedge
\Omega^{c}_{~b}-  \Omega^{a}_{~c} \wedge \omega^{c}_{~b}=0~,
\end{equation}
namely, $D\Omega^{a}_{~b}=0$.

In general relativity, the connection of space-time manifold,
$\theta^{a}_{~b}$, is determined in terms of the vierbeins and
inverse vierbeins and is related to the Christoffel symbol
$\Gamma^{\mu}_{\nu\rho}$, where $\Gamma^{\mu}_{\nu\rho}$ is
uniquely determined by two conditions, the covariant constancy of
the metric and the absence of torsion. In our framework, there is
not torsion-free condition. The connection $\omega^{a}_{~b}$ can
be determined by the condition that the covariant exterior
derivative of ${\cal R}^{a}_{~b}$ must be vanishing. That is to
say,
\begin{equation}
\label{cod-of-R22} D{\cal R}^{a}_{~b}=D(i_{X}\Omega^{a}_{~b})={\rm
d }\cdot i_{X}\Omega^{a}_{~b}+\omega^{a}_{~c} \wedge
i_{X}\Omega^{c}_{~b}+ i_{X}\Omega^{a}_{~c} \wedge
\omega^{c}_{~b}=0~.
\end{equation}
From the above equation, it is easy to prove that the complex
connection $\omega^{a}_{~b}$ must satisfy the following constraint
equation
\begin{equation}\label{xequ2}
\begin{array}{l}
\displaystyle\left(\omega^{a}_{bj,e,i}-\omega^{a}_{bi,e,j}-\omega^{a}_{be,j,i}
 + \omega^{a}_{be,i,j}+\omega^{a}_{ci}\omega^{c}_{bj,e}
-\omega^{a}_{cj}\omega^{c}_{bi,e}-\omega^{a}_{ci}\omega^{c}_{be,j}
\right.\\[0.4cm]
\displaystyle~~~~~~~~~~~\left.
+\omega^{a}_{cj}\omega^{c}_{be,i}+\omega^{a}_{ci,e}\omega^{c}_{bj}-\omega^{a}_{cj,e}\omega^{c}_{bi}
-\omega^{a}_{ce,i}\omega^{c}_{bj}+\omega^{a}_{ce,j}\omega^{c}_{bi}
\right)k^{i}\wedge k^{j}=0~,
\end{array}
\end{equation}
because $\omega^{a}_{ci,e}\equiv \partial_{e}\omega^{a}_{ci}$ and
$i_{X}\Omega^{a}_{~b}=i_{X}\cdot {\rm d } \omega^{a}_{~b}=
(\omega^{a}_{~bj,i}-\omega^{a}_{~bi,j})\gamma^{i} k^{j}$.

Conversely, the complex curvature $\Omega^{a}_{~b}$ can be
directly obtained from ${\cal R}^{a}_{~b}$ as follows
\begin{equation}
\label{inner} \Omega^{a}_{~b}=\frac{1}{2}\gamma_{i}k^{i}\wedge
{\cal R}^{a}_{~b} ~.
\end{equation}

In the following we will demonstrate that Einstein's gravitational
equation (\ref{daor1}) can be deduced from Eq.(\ref{daor21}) and
Eq.(\ref{xequ2}). Define the Hermitian operator 1-form
\begin{equation}
\label{daor23} \hat{W}\equiv i \delta^{a}_{~b}{\rm d }+ i
(\omega^{a}_{~b}+\lambda\gamma^{c}\Omega^{a}_{~bc})\wedge~,
\end{equation}
then Eq.(\ref{daor21}) is rewritten as $\hat{W}k^{b}=0$.
Multiplying both sides of Eq.(\ref{daor21}) by the operator
$\hat{W}$ yields
\begin{eqnarray}
\label{daor24} \nonumber 0&=&\hat{W} \hat{W} k^{b}
\\
\nonumber
 &=& \left[ {\rm d
}(\omega^{a}_{~b}+\lambda{\cal
R}^{a}_{~b})+(\omega^{a}_{~c}+\lambda{\cal R}^{a}_{~c})\wedge
(\omega^{c}_{~b}+\lambda{\cal R}^{c}_{~b}) \right] \wedge  k^{b}
\\
 &=& \left[ \Omega^{a}_{~b}+\lambda({\rm d
}{\cal R}^{a}_{~b}+\omega^{a}_{~c}\wedge {\cal R}^{c}_{~b}+{\cal
R}^{a}_{~c}\wedge\omega^{c}_{~b})+\lambda^{2}{\cal
R}^{a}_{~c}\wedge {\cal R}^{c}_{~b} \right] \wedge  k^{b}~.
\end{eqnarray}
Since we have required that $D{\cal R}^{a}_{~b}={\rm d } {\cal
R}^{a}_{~b}+\omega^{a}_{~c} \wedge {\cal R}^{c}_{~b}+ {\cal
R}^{a}_{~c} \wedge \omega^{c}_{~b}=0$ in Eq.(\ref{cod-of-R22}),
the above equation reduces to
\begin{eqnarray}
\label{reduce01}  \left( \Omega^{a}_{~b}+\lambda^{2}{\cal
R}^{a}_{~c}\wedge {\cal R}^{c}_{~b} \right) \wedge  k^{b}=0~.
\end{eqnarray}

A sufficient, but not necessary, solution for Eq.(\ref{reduce01})
is expressed as
\begin{equation}
\label{daor2601}\Omega^{a}_{~b}=-\lambda^{2}{\cal
R}^{a}_{~e}\wedge{\cal R}^{e}_{~b}=-\lambda^{2}
\gamma^{c}\gamma^{d}\Omega^{a}_{~ec}\wedge
\Omega^{e}_{~bd}=-\lambda^{2} \Omega^{a~c}_{~e}\wedge
\Omega^{e}_{~bc}~.
\end{equation}
Now we will demonstrate that the above solution is not
self-contradictory. Eq.(\ref{daor25}) and Eq.(\ref{cod-of-R22})
make sure that
\begin{eqnarray}
\label{cod-of-R-times} D({\cal R}^{a}_{~e}\wedge{\cal
R}^{e}_{~b})=(D{\cal R}^{a}_{~e})\wedge{\cal R}^{e}_{~b}-{\cal
R}^{a}_{~e}\wedge(D{\cal R}^{e}_{~b})=0~.
\end{eqnarray}
Hence Eq.(\ref{daor2601}) is consistent with Bianchi identities.

Componentwise Eq.(\ref{daor2601}) can be written as
\begin{eqnarray}
\nonumber
 \frac{1}{2}\Omega^{a}_{~bij}k^i\wedge k^j
&=&-\lambda^{2}
\gamma^{c}\gamma^{d}(\Omega^{a}_{~eci}k^i)\wedge(\Omega^{e}_{~bdj}k^j)
\\
\label{daor2602}
 &=&-\frac{1}{2}\lambda^{2}
(\Omega^{a~c}_{~e~i}\Omega^{e}_{~bcj}
-\Omega^{a~c}_{~e~j}\Omega^{e}_{~bci})k^i\wedge k^j~,
\end{eqnarray}
namely
\begin{eqnarray}
\label{daor2603}
 \Omega^{a}_{~bij} =\lambda^{2}
(\Omega^{a~c}_{~e~j}\Omega^{e}_{~bci}
-\Omega^{a~c}_{~e~i}\Omega^{e}_{~bcj})~.
\end{eqnarray}

It is well known, in gauge field theory the freedom of gauge field
is too much to describe the physical system. Theorists must
introduce some kind of gauge fixing condition, such as Coulomb
gauge in QED or Landau gauge in QCD. In our framework we propose
the following gauge fixing condition
\begin{eqnarray}
\label{gauge-fixing} \theta^{a}_{~c}\wedge
\tilde{B}^{c}_{~b}+\tilde{B}^{a}_{~c}\wedge\theta^{c}_{~b}=0~.
\end{eqnarray}
In this gauge the curvature $\Omega^{a}_{~b}$ reduces to
\begin{eqnarray}
\label{total-curv} \Omega^{a}_{~b}= R^{a}_{~b}+i\epsilon {\tilde
F}^{a}_{~b}~.
\end{eqnarray}

Inserting above equation into Eq.(\ref{daor2602}), we acquire two
equations which are respectively real part and complex part of an
equation. They are expressed as follows
\begin{eqnarray}
\label{component001}
 R^{a}_{~bij}
 =\lambda^{2}
 \left(R^{a~c}_{~e~j}R^{e}_{~bci}
-R^{a~c}_{~e~i}R^{e}_{~bcj} \right)
 + \lambda^{2}\epsilon^{2} \left({\tilde F}^{a~c}_{~e~i}{\tilde
F}^{e}_{~bcj} - {\tilde F}^{a~c}_{~e~j}{\tilde
F}^{e}_{~bci}\right)~,
\end{eqnarray}
and
\begin{eqnarray}
\label{component002}
 {\tilde F}^{a}_{~bij}
 =\lambda^{2}
 \left({\tilde F}^{a~c}_{~e~j}R^{e}_{~bci}
+R^{a~c}_{~e~j}{\tilde F}^{e}_{~bci} -  R^{a~c}_{~e~i}{\tilde
F}^{e}_{~bcj} -{\tilde F}^{a~c}_{~e~i}R^{e}_{~bcj}\right)~.
\end{eqnarray}
Contracting $a$ with $i$ in $R^{a}_{~bij}$ yields
\begin{eqnarray}\label{component003}
 R_{\mu\nu}
 =R^{i}_{~\mu i\nu}= \lambda^{2}
 \left( R^{i~c}_{~e~\nu}R^{e}_{~\mu c i}
 - R^{i~c}_{~e~i}R^{e}_{~\mu c \nu} \right)+ \lambda^{2}\epsilon^{2}
 \left( {\tilde F}^{i~c}_{~e~i}{\tilde F}^{e}_{~\mu
c\nu} -{\tilde F}^{i~c}_{~e~\nu}{\tilde F}^{e}_{~\mu c i}
\right)~,
\end{eqnarray}
where $ R_{\mu\nu}$ is called Ricci tensor.

Einstein's gravitational equation (\ref{daor1}) can be rewritten
as follows
\begin{equation}
\label{einre} R_{\mu\nu}=8\pi G \left(T_{\mu\nu}-\frac{1}{2} ~
g_{\mu\nu}T^{\rho}_{~\rho}\right)+\Lambda g_{\mu\nu}~,
\end{equation}
where $T^{\rho}_{~\rho}$ denotes the contraction of the
stress-energy tensor $T_{\mu\nu}$. In the case of SU(1,3) gauge
field, inserting Eq.(\ref{daor20}) into Eq.(\ref{einre}) yields
\begin{equation}
\label{eingauge} R_{\mu\nu}= G\left[\frac{1}{2}  g_{\mu\nu}{\rm
tr}({\tilde F}_{\alpha\beta}{\tilde F}^{\alpha\beta}) -2{\rm tr}(
{\tilde F}_{\mu}^{~\alpha} {\tilde F}_{\nu\alpha})\right]+\Lambda
g_{\mu\nu} ~.
\end{equation}

From Eq.(\ref{component003}), the Einstein tensor
$R_{\mu\nu}-\frac{1}{2}g_{\mu\nu}R^{\rho}_{~\rho}$ leads to
\begin{eqnarray}\nonumber
&&\lambda^{2}\epsilon^{2}\left[
 \left( {\tilde F}^{i~c}_{~e~i}{\tilde F}^{e}_{~\mu
c\nu} -{\tilde F}^{i~c}_{~e~\nu}{\tilde F}^{e}_{~\mu c i}
\right)-\frac{1}{2}g_{\mu\nu}\left({\tilde F}^{i~c}_{~e~i}{\tilde
F}^{e\rho}_{~~ c\rho} -{\tilde F}^{i~c}_{~e~\rho}{\tilde
F}^{e\rho}_{~~ c i}\right)\right]
\\
\label{contr} && =\lambda^{2}\epsilon^{2}\left[
  {\tilde F}^{i~c}_{~e~i}{\tilde F}^{e}_{~\mu
c\nu} -{\rm tr}\left({\tilde F}^{c}_{~\nu}{\tilde F}_{\mu c}
\right)-\frac{1}{2}g_{\mu\nu}{\tilde F}^{i~c}_{~e~i}{\tilde
F}^{e\rho}_{~~c\rho} +\frac{1}{2}g_{\mu\nu}{\rm tr}\left({\tilde
F}^{c}_{~\rho}{\tilde F}^{\rho}_{~c}\right)\right] ~.
\end{eqnarray}
We have indicated that the energy-momentum tensor of SU(1,3) gauge
field should be included in the total energy-momentum tensor in
Einstein's gravitational equation. Comparing Eq.(\ref{contr}) with
Eq.(\ref{daor1}), Eq.(\ref{daor20}) directly yields a simple
identity
\begin{eqnarray}
\label{graconstant} G= \lambda^{2}\epsilon^{2} ~,
\end{eqnarray}
where the Newtonian gravitational constant $G$, the coupling
constant of gauge field $\epsilon$ and the coupling constant of
the daor field $\lambda$ are connected.

We can also draw a conclusion that the term $\lambda^{2}
\left(R^{i~c}_{~e~\nu}R^{e}_{~\mu c i} - R^{i~c}_{~e~i}R^{e}_{~\mu
c \nu}\right)$ in Eq.(\ref{component003}) plays the same role as
that of the term $\Lambda g_{\mu\nu}$ in Eq.(\ref{eingauge}) by
comparing Eq.(\ref{component003}) with Eq.(\ref{eingauge}), namely
\begin{eqnarray}
\label{cosconstant}  \lambda^{2} \left(R^{i~c}_{~e~\nu}R^{e}_{~\mu
c i} - R^{i~c}_{~e~i}R^{e}_{~\mu c \nu}\right) \Longrightarrow
\Lambda g_{\mu\nu}~.
\end{eqnarray}

In this section we have proposed the first-order nonlinear daor
field equation (\ref{daor21}), from which Einstein's gravitational
equation (\ref{einre}) can be deduced. We have argued that the
cosmological constant term should be substituted by a Ricci
squared term. This fact will be studied further in the next
section.


\section{Dark Energy Originates from the Self-coupling
of the Space-time Curvature \label{sec:Sec6}}
\renewcommand{\theequation}{6.\arabic{equation}}
\setcounter{equation}{0}

Recent cosmological
observations~\cite{sch98,rie98,per99,ber00,jaf01,ben03} have not
only strengthened and expanded the big bang model, but they have
also revealed some surprises. In particular, most of the universe
seems to be made of something fundamentally different from the
matter of which we are made. About 23$\%$ of the total energy is
dark matter, composed of particles most likely formed early in the
Universe. About seventy percents is in a smooth dark energy whose
gravitational effects began causing the expansion of the Universe
to speed up just a few billion years ago. However, the nature of
this energy and, therefore, the meaning of the non-zero
cosmological constant which is needed in the equations that
describe the nowaday acceleration of the expansion is still a
mystery.

Recently there have been a number of different
attempts~\cite{car04} to modify gravity to yield accelerating
universe at late times. Specially the models~\cite{car05} in which
the higher order curvature invariants is directly added to the
Einstein-Hilbert action have been widely investigated. The
cosmological models in Ricci squared gravity has also been studied
in Ref.\cite{abf04}. In last section we have showed that in our
scenario of the daor field, the cosmological constant term in
Einstein's equation must be substituted by the term that describes
the effect of the self-coupling of the space-time curvature. It is
well known that the notion of dark energy is an evolutionary
expansion of the cosmological constant. Hence, from this point of
view, one can say that dark energy originates from the
self-coupling of the space-time curvature in our scenario.

To demonstrate the physical meaning of Eq.(\ref{cosconstant})
explicitly, we consider a simple case that the space-time is a
maximally symmetric manifold. The mathematical theory of symmetric
space is elaborate and has been used extensively in modern
physics. Anti de Sitter, de Sitter and Minkowskian space-times are
all maximally symmetric space. The Maldacena conjecture on the
AdS/CFT correspondence has become an important part of string
theory~\cite{mal98}. The curvature tensor of a four dimensional
maximally symmetric space-time is given by
\begin{eqnarray}
\label{maxRicci}
R_{\sigma\rho\alpha\nu}=\frac{R^{\lambda}_{~\lambda}}{12}\left\{
g_{\nu\rho}g_{\sigma\alpha}-g_{\alpha\rho}g_{\sigma\nu} \right\}
~.
\end{eqnarray}
The Ricci tensor is easily acquired by contracting the suffix
$\sigma$ with the suffix $\alpha$, that is
\begin{eqnarray}
\label{Ricci}
R_{\mu\nu}=\frac{R^{\lambda}_{~\lambda}}{4}g_{\mu\nu} ~.
\end{eqnarray}
It is well known that $R^{\lambda}_{~\lambda}$ is a constant in
the case of maximally symmetric space-time, namely
$R^{\lambda}_{~\lambda}=R$, $R$ being the curvature of the
space-time~\cite{wei72}.

Inserting Eq.(\ref{maxRicci}) into Eq.(\ref{cosconstant}) yields
\begin{eqnarray}
\label{611} \frac{\lambda^{2}}{24}R^{2} g_{\mu\nu}\Longrightarrow
\Lambda g_{\mu\nu}~.
\end{eqnarray}
then the above equation reduces to
\begin{eqnarray}
\label{cosconstant++}  \Lambda =\frac{\lambda^2}{24}R^{2} ~.
\end{eqnarray}
Similar to the definition of the vacuum energy density,
$<\rho>=\Lambda/(8\pi G)$, we can define the dark energy density
as follows
\begin{eqnarray}
\label{vacdensity} \rho=\frac{\lambda^2}{24}\frac{R^{2}}{8\pi
G}=\frac{R^{2}}{192\pi \epsilon^{2}}~,
\end{eqnarray}
the above equation exactly sets up a relationship between the dark
energy density and the curvature of maximally symmetric
space-time. Hence the dark energy parameter $\Omega_{\Lambda}$ is
given
\begin{eqnarray}
\label{vac++} \Omega_{\Lambda}=\frac{8\pi G
\rho}{3H^{2}_{0}}=\frac{\lambda^2 R^{2}}{72H^{2}_{0}}~.
\end{eqnarray}
We can express the dark energy density in terms of a mass scale,
\begin{eqnarray}
\label{vacdensity++} \rho=M^{4}_{de}~,
\end{eqnarray}
so the updated observational result is
\begin{eqnarray}
\label{vacden}M_{de} \sim 10^{-3} {\rm eV}~.
\end{eqnarray}
In our scenario the coupling constant $\epsilon$ is the unique
coupling constant of all gauge fields, so the mass scale of
$\frac{1}{\epsilon}$ should be chosen as $M_{GUT}\simeq 2\times
10^{16}{\rm GeV}$. The main motivation for this selection is that,
at least in supersymmetric models, the running gauge couplings of
the standard model unify at the scale $M_{GUT}$~\cite{ekn90},
hinting at the presence of a grand unified theory involving a
higher symmetry with a single gauge coupling. Therefore, we can
get the curvature of maximally symmetric space-time, that is
\begin{eqnarray}
\label{curv} R\simeq\frac{M^{2}_{de}}{M_{GUT}}\simeq 1\times
10^{-23} ~{\rm m}^{-1}~.
\end{eqnarray}
What we obtained here is in agreement with the result of recent
astronomical
observations~\cite{sch98,rie98,per99,ber00,jaf01,ben03}.

After A.H.~Guth, it is widely believed that the horizon problem
and the flatness problem in modern cosmology can be solved once we
accept that our universe underwent an inflationary evolution in
the early epoch~\cite{gut81,lr99}. The recent observations on the
CMB radiation~\cite{ber00,jaf01,ben03} confirmed the existence of
the early inflationary cosmological epoch as well as the
accelerated expansion of the present universe. The cosmological
observations~\cite{ber00,jaf01,ben03} also demonstrated that the
large scale structures in present university possibly originate
from the quantum fluctuation in the epoch of
inflation~\cite{bar80}. In our scenario, dark energy originates
from the self-coupling of the space-time curvature. Obviously
\begin{eqnarray}
\label{inflation}  \lambda^{2} \left(R^{i~c}_{~e~\nu}R^{e}_{~\mu c
i} - R^{i~c}_{~e~i}R^{e}_{~\mu c \nu}\right) \propto R^2~.
\end{eqnarray}
This equation indicates that our scenario maybe has provided a
natural mechanism for the early inflation when the curvature of
our universe is very large. This mechanism will have no
difficulties on the exit of inflation which has been confronted by
many inflationary scenarios. This topic will be explicitly
discussed in our forthcoming papers.

\section{Conclusion \label{sec:Sec7}}
\renewcommand{\theequation}{6.\arabic{equation}}
\setcounter{equation}{0}

The locally complexified vierbein (or tetrad) field has been
discussed. We renamed it and suggested calling it the daor field.
Under this expanded symmetry, we introduced the daor geometry
where the connection is complex. After setting up the geometric
tools, we proposed a first-order nonlinear equation, from which
Einstein's gravitational equation can be deduced. The real part of
the daor field coupling equation can also be regarded as
Einstein's equation endowed with the cosmological constant term.
Our equation has shown that dark energy originates from the
self-coupling of the space-time curvature. The energy-momentum
tensor of dark energy is affected by the scale of the grand
unified theory. The dark energy density obtained in our scenario
is in the same order as that given by the astronomical
observations.

 {\bf Acknowledgement:}  I would like to thank Prof. D. V. Ahluwalia-Khalilova, Prof. L. Z. Fang, Prof. B.-Q.
Ma, Prof. D. S. Du, Prof. C. J. Zhu, Prof. Z. Chang, Long-Mei Zhu
and You-Ping Dai for their help and encouragement.

\end{document}